\documentclass[%
 reprint,
 superscriptaddress,
 amsmath,amssymb,
 aps,
]{revtex4-1}
\usepackage{braket}
\usepackage{graphicx}% Include figure files
\usepackage{dcolumn}% Align table columns on decimal point
\usepackage{bm}% bold math
\usepackage{float}
\usepackage[usenames,dvipsnames]{color}
\usepackage[colorlinks=true, citecolor=blue, linkcolor=blue, urlcolor=blue]{hyperref}
\usepackage{bm}
\usepackage{cleveref} %to use \cref command
\usepackage{multirow}

\newcommand{\pars}[1]{{(#1)}}

\newcommand{\HSYK}{\hat{H}_{\textsc{syk}}}
\newcommand{\Npar}{N_{\mathrm{par}}}

\begin{document}

\title{Can neural quantum states learn volume-law ground states?}

\newcommand{\affiliationRWTH}{
Institut f\"ur Theorie der Statistischen Physik, RWTH Aachen University and JARA-Fundamentals of Future Information Technology, 52056 Aachen, Germany
}
\newcommand{\affiliationMPSD}{
Max Planck Institute for the Structure and Dynamics of Matter,
Center for Free-Electron Laser Science (CFEL),
Luruper Chaussee 149, 22761 Hamburg, Germany
}

\newcommand{\affiliationBristol}{
H~H~Wills Physics Laboratory, University of Bristol, Bristol BS8 1TL, United
Kingdom 
}

\author{Giacomo Passetti}
\affiliation{\affiliationRWTH}

\author{Damian Hofmann}
\affiliation{\affiliationMPSD}

\author{Pit Neitemeier}
\affiliation{\affiliationRWTH}

\author{Lukas~Grunwald}
\affiliation{\affiliationRWTH}
\affiliation{\affiliationMPSD}

\author{Michael~A. Sentef}
\affiliation{\affiliationBristol}
\affiliation{\affiliationMPSD}

\author{Dante M.~Kennes}
\affiliation{\affiliationRWTH}
\affiliation{\affiliationMPSD}

\date{\today}
\begin{abstract}
We study whether neural quantum states based on multi-layer feed-forward networks can find ground states which exhibit volume-law entanglement entropy. As a testbed, we employ the paradigmatic Sachdev-Ye-Kitaev model.
We find that both shallow and deep feed-forward networks require an exponential number of parameters in order to represent the ground state of this model.
This demonstrates that sufficiently complicated quantum states, although being physical solutions to relevant models and not pathological cases, can still be difficult to learn to the point of intractability at larger system sizes. This highlights the importance of further investigations into the physical properties of quantum states amenable to an efficient neural representation.
\end{abstract}
\maketitle

%%%%%%%%%%%%%%%%%%%%%%%%%%%%%%%%%%%%%%%%%%%%%%%%%%%%%%%%%%%%%%%%%%%%%% INTRODUCTION
%%%%%%%%%%%%%%%%%%%%%%%%%%%%%%%%%%%%%%%%%%%%%%%%%%%%%%%%%%%%%%%%%%%%%%

\paragraph*{Introduction.---}
The exponential complexity of representing general quantum many-body states is a key challenge in computational quantum physics.
To simulate systems beyond small sizes tractable by exact diagonalization methods, it is necessary to find an efficient representation of quantum states of interest.
This is made possible by the fact that physically relevant states usually possess a high degree of structure, compared to an arbitrary Hilbert space vector.
As a prominent example, ground states of local, gapped Hamiltonians exhibit an area law of the entanglement entropy, i.e., an entanglement entropy that scales like the boundary of the subregion instead of its volume.
For systems with a low dimensionality, typically 1D, the area law allows for an efficient representation of the wave function as a matrix product state, which can be simulated by algorithms such as the density matrix renormalization group (DMRG) \cite{Verstraete2006,Verstraete2008,Eisert2010,Schollwck2011,Cirac2021}.

However, many quantum states of physical interest display a volume law scaling of the entanglement entropy \cite{Bianchi2022},
for which generally applicable efficient representations are not known to this date.
One class of variational approximations that has been studied to overcome this challenge are neural quantum states (NQS) \cite{Carleo2017}, which are based on an artificial-neural-network representation of the wave function's probability amplitudes \cite{Schmidhuber2015,LeCun2015,Goodfellow2016} and have shown promising results for the study of discrete lattice models even beyond one dimension \cite{Glasser2018,Clark2018,Kaubruegger2018,Choo2019,Fabiani2019,Schmitt2020,Fabiani2021,Astrakhantsev2021,Roth2022}.
Notably, it has been shown that a shallow NQS ansatz is able to efficiently represent quantum states featuring volume-law entanglement \cite{Deng2017, Sun2022}, suggesting that this method could complement tensor network techniques for the purpose of uncovering the physics of highly entangled states.
Nevertheless, while for matrix product states and more general tensor-network-based approaches it is known how the entanglement scaling limits the representation capabilities of the ansatz \cite{Eisert2010},
there is so far no analogous physical property that directly relates to the ability of an NQS to learn a given quantum state.
Universal approximation theorems, which have been proven for several broad classes of neural networks, guarantee that, in the limit of infinite network size, a neural network ansatz can theoretically represent any continuous function to arbitrary precision \cite{Cybenko1989,Hornik1991,Pinkus1999,Kidger20a}.
Still, these results do not provide bounds on the scaling of the required number of parameters with the system size.
For practical applications of NQS, it is thus a central question to determine which classes of quantum many-body states can be efficiently represented that are impossible to tackle with other established variational ans\"atze.

In this Letter, we investigate the capabilities of NQS based on shallow and deep feed-forward neural networks (FFNNs) to represent ground states of the Sachdev-Ye-Kitaev (SYK) model \cite{Sachdev1993,kitaevSimpleModel2015,maldacenaRemarksSachdevYeKitaev2016}, which is a paradigmatic model for quantum chaos and non-Fermi liquid behavior \cite{chowdhurySachdevYeKitaevModels2022} and which features a volume-law entanglement in the ground state \cite{Liu2018}.
We present a systematic study of the representation accuracy achieved by the FFNN in dependence of the network hyperparameters.
We find an exponential dependence on the system size for the number of network parameters required to learn the SYK ground state.
This demonstrates limitations of fully general NQS to learn complicated quantum ground states of physical interest.

\paragraph*{Model.---}

\begin{figure}[tbp]
\centering
\minipage{0.45\textwidth}
  \includegraphics[width=0.8\linewidth]{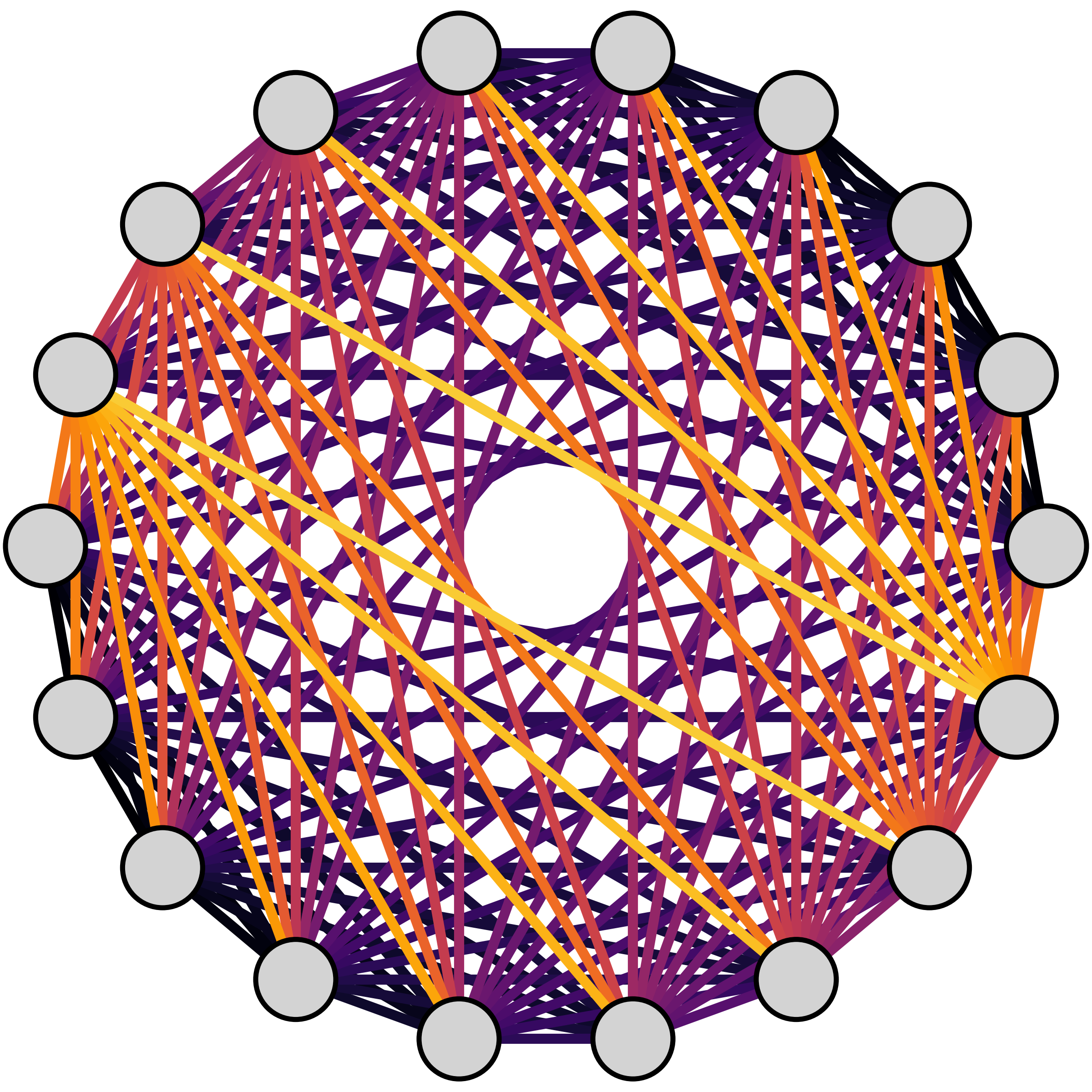}
\endminipage\hfill
\caption{Cartoon representation of the SYK model.
Gray circles represent lattice sites and  every different colour shown has two  corresponding lines in total connecting four sites. Each color represents one element of the coupling matrix $J_{ij;kl}$ of the SYK model defined by Eq. \eqref{eq:def_syk}.}
\label{fig:scaling-number_of_parameters.}
\end{figure}

The SYK model describes strongly correlated fermions on $L$ sites and is defined by the Hamiltonian \cite{Sachdev1993,kitaevSimpleModel2015,maldacenaRemarksSachdevYeKitaev2016}
\begin{equation}
    \label{eq:def_syk}
    \HSYK\left( J \right) = \frac{1}{(2L)^{3/2}} \sum_{ijkl} J_{ij;kl} \, \hat c^\dagger_i \hat c^\dagger_j \hat c_k \hat c_l,
\end{equation}
where $\hat c_i^{(\dagger)}$, $i \in \{1, \dots L\},$ are fermionic ladder operators. The vertices $J_{ij;kl}$ have the symmetry $J^*_{ij;kl} = J_{lk;ji}$ and $J_{ij;kl} = -J_{ji;kl}$ and are random, uncorrelated, all-to-all couplings that are drawn from a Gaussian unitary ensemble (GUE) \cite{Akemann2015} with mean $\mathbb{E}\left[J_{ij;kl} \right] = 0$ and variance $\mathbb{E}\left[\vert J_{ij;kl}\vert^2\right] = 1$ \cite{chowdhurySachdevYeKitaevModels2022}.
Consequently, quantities of physical interest are expectation values over the ensemble of couplings~$J$, which is evaluated after the quantum-expectation value. The ground state of the SYK model describes a strongly correlated non-Fermi liquid without quasi-particle excitations \cite{chowdhurySachdevYeKitaevModels2022}, that exhibits volume-law entanglement entropy \cite{Sachdev2016,zhangQuantumEntanglement2022}.
In the thermodynamic limit the model becomes self-averaging and exactly solvable, but despite this exact solvability, the ground state is not a Gaussian state, i.e.~not a product of single particle wave-functions \cite{Haldar2021}.
At finite sizes, particularly studied in the context of quantum chaos \cite{altlandQuantumErgodicity2018,PhysRevD.94.126010} and experimental realizations \cite{brzezinskaEngineeringSYK2022}, no exact solutions are known.
Different variational ans\"atze to represent the ground state have been proposed recently \cite{Haldar2021, Kim2021}.
Here the model can be analyzed by employing approximations, or numerically, by drawing a set of couplings $\{J^{(n)}\}_{n=1}^N$ from the GUE, constructing the corresponding Hamiltonians $\HSYK(J^{(n)}),$ and solving for the ground states $\ket{\Psi_{\textsc{GS}}(J^{(n)})}$. 
Finally, the properties of interest, such as expectation values, are averaged over this ground state ensemble.
Because of the self-averaging property of the SYK model, it suffices to evaluate expectation values for a single realization of $J$ in the thermodynamic limit \cite{chowdhurySachdevYeKitaevModels2022}.

\begin{figure}[tbp]
\centering
\minipage{0.45\textwidth}
  \includegraphics[width=\linewidth]{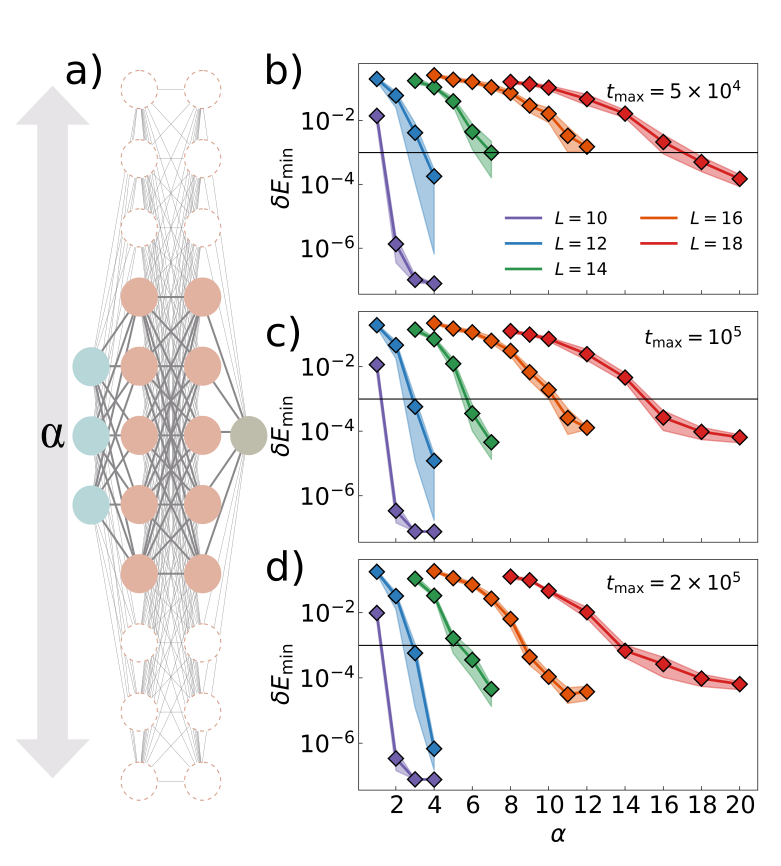}
\endminipage
\caption{(a) Shallow fully-connected feed-forward neural network,  $\alpha$ denotes the hidden unit density of each layer and thus parametrizes the width of the network.  (b), (c), (d) Relative ground state energy error $\delta E$ as function of the network width $\alpha$ for several system sizes and random initializations after (b) $5 \times 10^{4}$, (c) $10^{5} $, and (d) $2\times 10^{5}$ simulation steps, respectively.
The color of each set of data points corresponds to the average over four independent 
realizations of the network initial weights, for the system size $L$ as indicated in the legend. The coloured areas give the maximum and minimum values of $\delta E$ for the four independent runs.
Black bars indicate $\delta E_{\rm threshold}=10^{-3}$.}
\label{fig:scaling_alpha}
\end{figure}

\begin{figure}[tbp]
\centering
\minipage{0.45\textwidth}
  \includegraphics[width=\linewidth]{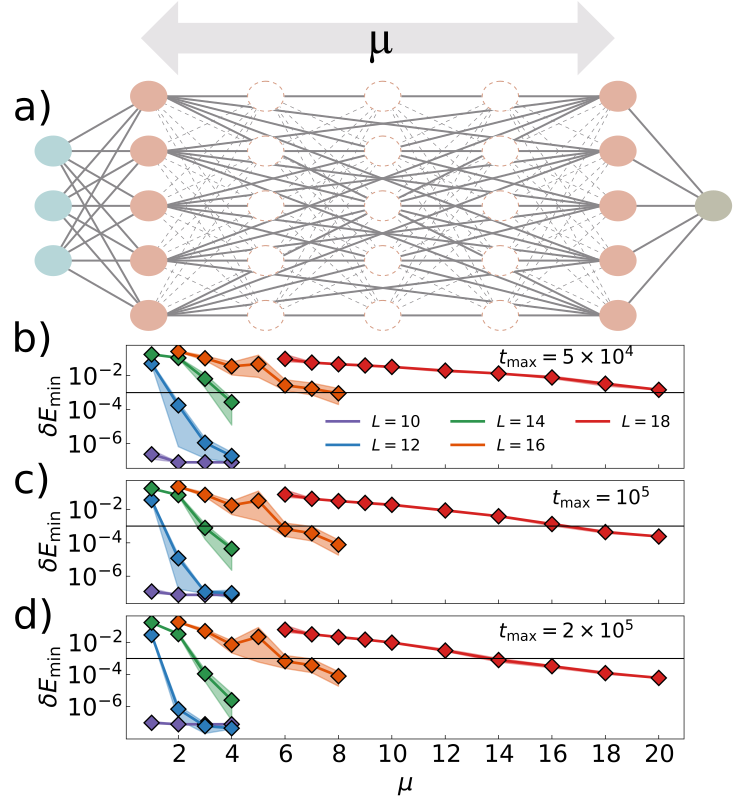}
\endminipage
\caption{(a) Deep fully-connected feed forward neural network, $\mu$ denotes the number of layers and thus the network depth.   (b), (c), (d) Relative energy error $\delta E$ as function of the network number of layers $\mu$ for several system sizes and random initializations after (b) $5 \times 10^{4}$, (c) $10^{5} $, and (d) $2\times 10^{5}$ simulation steps, respectively.
The color of each set of data points corresponds to the average over four independent 
realizations of the network's initial weights, for the system size $L$ as indicated in the legend. The coloured areas give the maximum and minimum values of $\delta E$ for the four independent runs.
Black bars indicate $\delta E_{\rm threshold}=10^{-3}$.}
\label{fig:scaling_mu}
\end{figure}

\paragraph*{Network architecture.---}
We use a fully-connected FFNN [Figs.~\ref{fig:scaling_alpha}(a), \ref{fig:scaling_mu}(a)]
\begin{align}
\label{eq:ffnn}
\begin{split}
    F(x) &= f^\pars{\mu} \circ \cdots \circ f^\pars{1} (x), \\
    f^\pars{l}(y) &= \phi(W^\pars{l} y + b^\pars{l})
\end{split}
\end{align}
which is a composition of $\mu$ layers $f^\pars{l},$ each applying an affine transformation and a scaled exponential linear unit (SELU) activation function~$\phi$ \cite{Klambauer2017} as pointwise nonlinearity.
Each layer has $\alpha L$ neurons, where $\alpha$ is the fixed hidden unit density.
The output of the final layer is reduced to a (scalar) log-probability amplitude with respect to the computational basis $\{\ket{x}\}$ by an exponential sum,
\begin{align}
\label{eq:nqs-wf}
    \log\braket{x | \psi_\theta} = \log\sum_{i=1}^{\alpha L} \exp[{F_i(x)}].
\end{align}
Here, $\theta$ denotes the vector of all variational parameters, which contains all entries of the weight matrices $W^\pars{l}$ and bias vectors $b^\pars{l}.$
The variational parameters and therefore network outputs are complex numbers, with the activation function being applied separately to real and imaginary parts.
The total number of network parameters scales as $\Npar = \mathcal O(\mu\, \alpha^2 L^2)$.
We choose the occupation number basis (as has been done in previous NQS studies of fermionic molecular Hamiltonians \cite{Choo_2020,Yang2020,Hermann2022})
at half filling, which fixes the fermion number to $L/2.$
Therefore, the input to the neural network \eqref{eq:ffnn} is a vector of occupation numbers $x \in \{0, 1\}^L$ such that $\sum_i x_i = L/2.$

We have verified our results for several variations of this network architecture.
In particular, we have evaluated using $\tanh$ as nonlinear activation function as well as the addition of skip connections, which can be used to counteract the increased training complexity of networks beyond a certain depth \cite{He2016,Li2018b}.
These variations did not achieve better results compared to those presented in the main text.
Details can be found in Section III of the supplemental material (SM) \cite{Supplement}.

\paragraph*{Optimization.---}
The ground state of the network is obtained by numerically minimizing the overlap difference
\begin{align}
\label{eq:energy}
    \delta O(\theta, J) = 1 - \left|\frac{\braket{\psi_{\theta}|\psi_{\rm GS}(J)}}{\braket{\psi_{\theta}|\psi_{\theta}}} \right|
\end{align}
between the variational state $\ket{\psi_\theta}$ and the ground state $\ket{\psi_{\rm GS}(J)}$
with respect to the variational parameters $\theta$ using Adam \cite{Kingma2014}.
We work with system sizes up to $L = 18$ sites, which are accessible via exact diagonalization (ED) and thus enable training using a supervised learning (SL) protocol targeting the overlap with the ED ground state $\ket{\psi_{\rm GS}(J)}$ \cite{Jonsson2018}.
The system size allows us to evaluate the loss function~\eqref{eq:energy} by summation over the full Hilbert space (preventing any potential errors arising from Monte Carlo sampling)
and to assess the quality of our results using the relative energy error
\begin{align}
\label{eq:energy-error}
    \delta E(\theta; J) = \frac{E(\theta; J) - E_\mathrm{GS}(J)}{E_\mathrm{GS}(J)}
\end{align}
compared to the target ground state energy $E_\mathrm{GS}(J) = \braket{\psi_{\rm GS}(J)|\HSYK(J)|\psi_{\rm GS}(J)}.$
Details on the optimization scheme are reported in Section II of the SM \cite{Supplement}.


\paragraph*{Results.---}
To start, we discuss the minimum energy error $\delta E_{\rm min} = {\rm min}|_{t\in [0, t_{\rm max}]}\delta E(\theta, J)$ reached within a maximum number of iterations $t_{\rm max}$ of the optimization protocol.
Figure~\ref{fig:scaling_alpha}(b) shows the dependence of $\delta E_{\rm min}$ on the network width $\alpha$ for a network with a fixed number of $\mu = 2$ layers, while Fig.~\ref{fig:scaling_mu}(b) shows the results as a function of network depth $\mu$ for deep networks with constant width $\alpha = 4.$
We select $\delta E_{\rm threshold} = 10^{-3}$ as a threshold error to assess successful convergence to the desired ground state.
With this threshold, one can see in Figs.~\ref{fig:scaling_alpha}(b) and \ref{fig:scaling_mu}(b) that at any fixed number of training iterations $t_{max}$ there is a systematic improvement of the accuracy with respect to increasing both $\alpha$ and $\mu$, as one would expect given the increased representation capabilities of the network at larger sizes.

\begin{figure}[tbp]
\centering
\minipage{0.45\textwidth}
  \includegraphics[width=\linewidth]{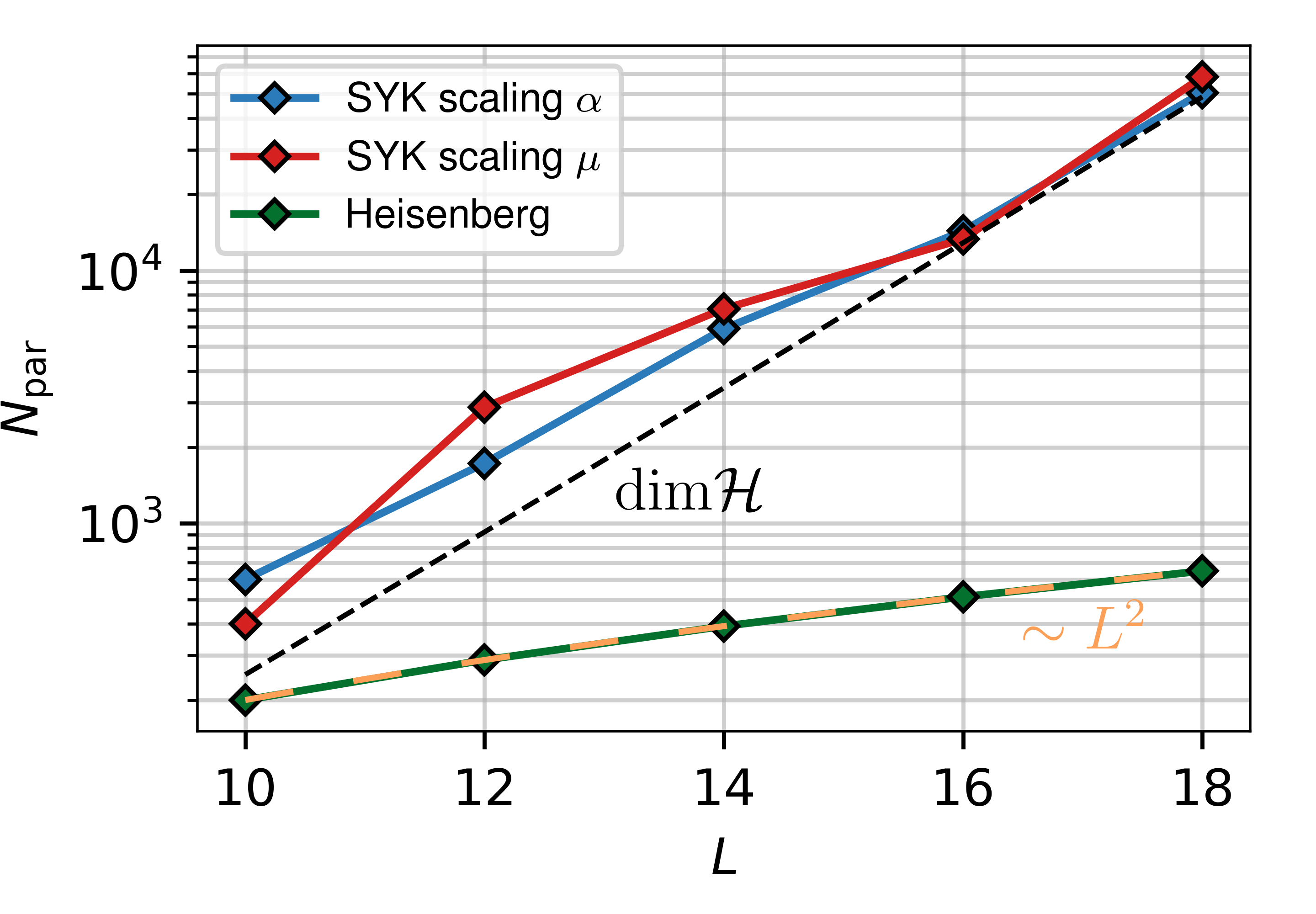}
\endminipage\hfill
\caption{Minimum number of parameters $N_{\rm par}$ required for the FFNN to learn the ground state of the SYK model as function of the system size $L$.
Results are shown for the scaling with network width in a shallow ($\mu = 2$) network (blue lines) and for the scaling with network depth for fixed $\alpha = 4$ (red line).
In both cases, an exponential scaling in the system size is observed, which matches the scaling of the full Hilbert space dimension $\dim \mathcal H$ (dashed line).
The $N_{\rm par}$ scaling for the ground state of the Heisenberg model (blue) and the associated quadratic polynomial law are reported for comparison.
}
\label{fig:scaling-number_of_parameters}
\end{figure}

Next, we determine the minimum number of variational parameters at which the network is able to learn the ground state with the desired energy of $\delta E_{\rm threshold}$.
Especially for the smallest system sizes, there is a clear transition between regimes where the network is able or unable to learn the state (in particular as a function of $\alpha$ in the shallow network).
For larger system sizes, it is somewhat more difficult to assess convergence.
While both very small and very large networks converge to energies above or below the desired threshold within a reasonable optimization time, there is an intermediate regime where the energy gets close to the threshold but only converges at very long time scales.
In order to systematically identify a value of $\alpha$ or $\mu$ at that boundary,
we have developed a criterion used to truncate optimization runs after a reasonable optimization time when those runs are predicted to ultimately converge to a $\delta E(\theta, J)$ higher than $\delta E_{\rm threshold}.$
See Section II B of the SM \cite{Supplement} for details.
 In Fig.~\ref{fig:scaling-number_of_parameters} we show the number of network parameters at the critical $\alpha_{\rm min}$ or $\mu_{\rm min}$ at which the network is able to reach the target energy accuracy threshold.
This allows for a comparison of network expressiveness for both varying width and depth on equal footing.
We find that for both the shallow and deep network, an exponentially growing number of parameters is needed to achieve the target energy error.
A comparison with the Hilbert space dimension reveals that the network only reaches this threshold once the number of variational parameters exceeds the number of probability amplitudes contained in the respective state vector.
Hence, we find that our deep feed-forward NQS ansatz as trained here does not learn a more efficient representation of the SYK ground state than the full state vector representation.
It is conceivable, in particular given the fully-connected nature of our ansatz, that there is some redundancy in the learned variational parameters, which could be used to achieve a degree of compression after training.
In order to investigate this possibility, we have performed a low-rank approximation based on singular value decomposition of the weight matrices \cite{Xue2013}, the details of which are reported in Section V of the SM \cite{Supplement}.
This analysis, however, has not revealed such an redundancy.

Our scaling results cannot be interpreted as an immediate consequence of the entanglement scaling of the SYK model, as NQS are known to be able to efficiently represent some volume-law quantum states \cite{Deng2017}, while they seem to fail for others (as shown here).
While a particular realization of the SYK Hamiltonian is of significantly higher complexity than a low-dimensional local lattice Hamiltonian (both because of its fully connected structure and the $\propto L^4$ randomly drawn interaction matrix elements), its ground state still exhibits more structure than a random Hilbert space vector.
Since it is well known that deep (and, in fact, already two-layer) networks are able to memorize even completely random data once the number of network parameters exceeds the number of data points \cite{Zhang2021},
these results provide evidence that our FFNN ansatz does not learn to utilize any of this structure but only manages to learn it as unstructured random data.
This is in stark contrast to more structured lattice Hamiltonians, where it is clear from previous works that neural quantum states can approximate ground state energies with sub-exponential scaling and thus do manage to make use of structure present in the quantum ground state \cite{Sehayek2019,Viteritti2022}, although exponential scaling results as a function of real time have been previously found for time-evolved states in a one-dimensional lattice spin model \cite{Lin2022}.
We have found comparable sub-exponential behavior when evaluating our training procedure on the ground state of the Heisenberg spin model
\(
    \hat H_{\rm Heisb} = \sum_{i=1}^N \sum_{q=1}^3 \hat\sigma^\pars{q}_i \hat\sigma^\pars{q}_{i + 1}
\)
on a one-dimensional chain with periodic boundary conditions
diagonalized in the same zero-magnetization subspace used for the SYK computations.
The scaling of the required number of parameters to reach $\delta E_{\rm threshold}$ in this model is also reported in Fig.~\ref{fig:scaling-number_of_parameters}.
In this case, a relatively small and fixed $\alpha = 1$ and $\mu = 2$ independent of the system size are sufficient to reach this threshold,
implying a polynomial scaling of the required number of parameters $N_{\rm par} = \mathcal{O}(L^{2}).$
This corresponds to an effective compression of the information contained in the exact state vector and allows to study sizes beyond those tractable by full state simulation \cite{Carleo2017,Sehayek2019}. However, the same approach fails to be useful in the more complex SYK model case.

\paragraph*{Discussion.---}
We have tackled the prototypical SYK model using an NQS variational ansatz, presenting a systematic study of the ability of deep FFNNs to learn the volume-law entangled ground states of this model.
Focusing on the scaling of the required number of parameters to describe the ground state to a desired and fixed accuracy
we find that the size of the FFNN ansatz needs to grow exponentially in the system size.
With this we show explicitly that the neural network ansatz is unable to efficiently represent SYK ground states in larger systems in spite of general results raising such hopes.
We have performed this analysis using a variety of training techniques (as detailed in the SM \cite{Supplement}), showing that the observed scaling is robust to such implementation choices.
While the proven capability of random RBMs to represent volume-law quantum states \cite{Deng2017,Sun2022} indicates that NQS methods have the potential to tackle problems out of the reach of established tensor-network based methods, our results demonstrate that the entanglement entropy is not the property that determines whether or not a physical quantum state can be efficiently represented by an NQS.
It remains an intriguing open question which other properties of a physical quantum state determine the efficient applicability of NQS-based methods.
NQS ans\"atze more specifically tailored to fermionic systems could potentially achieve better scaling \cite{Robledo2022,Hermann2022}.
Studies in this direction would help elucidate to what extent the nonlocal parity structure inherent to fermionic models \cite{Verstraete2005} affects the learnability of the SYK ground state.
Separating this influence from other sources of complexity, such as the lack of spatial structure and the disorder induced by random couplings, and thereby exploring the intermediate region between states that can be learned with compression (such as in the Heisenberg and similar spin models) and states that cannot (such as the SYK results presented here) can provide an improved understanding of the complexity of physical quantum states.

\begin{acknowledgements}
We acknowledge helpful discussions with Giuseppe Carleo, Sebastian Goldt, Javed Lindner, Claudia Merger, Alexandre René,  and Attila Szab\'o.
NQS calculations have been performed using \textsc{NetKet}~3 \cite{NetKet3,NetKet2} with \textsc{jax} \cite{Jax2018}.
Computations were performed on the HPC system Ada at the Max Planck Computing and Data Facility (MPCDF).
The authors also gratefully acknowledge computing time granted by the JARA Vergabegremium and provided on the JARA partition part of the supercomputer JURECA at Forschungszentrum Jülich \cite{Thrnig2021} under the project ID enhancerg.
We acknowledge support by the Max Planck-New York City Center for Nonequilibrium Quantum Phenomena.
\end{acknowledgements}

\bibliographystyle{apsrev4-1}
\bibliography{bibliography.bib}

\end{document}

\title{Supplementary Material for:\\Can neural quantum states learn volume-law ground states?}

\newcommand{\affiliationRWTH}{
Institut f\"ur Theorie der Statistischen Physik, RWTH Aachen University and JARA-Fundamentals of Future Information Technology, 52056 Aachen, Germany
}
\newcommand{\affiliationMPSD}{
Max Planck Institute for the Structure and Dynamics of Matter,
Center for Free-Electron Laser Science (CFEL),
Luruper Chaussee 149, 22761 Hamburg, Germany
}

\newcommand{\affiliationBristol}{
H~H~Wills Physics Laboratory, University of Bristol, Bristol BS8 1TL, United
Kingdom 
}

\author{Giacomo Passetti}
\affiliation{\affiliationRWTH}

\author{Damian Hofmann}
\affiliation{\affiliationMPSD}

\author{Pit Neitemeier}
\affiliation{\affiliationRWTH}

\author{Lukas~Grunwald}
\affiliation{\affiliationRWTH}
\affiliation{\affiliationMPSD}

\author{Michael~A. Sentef}
\affiliation{\affiliationBristol}
\affiliation{\affiliationMPSD}

\author{Dante M.~Kennes}
\affiliation{\affiliationRWTH}
\affiliation{\affiliationMPSD}

\maketitle
\onecolumngrid

\renewcommand\theequation{S\arabic{equation}}
\renewcommand\thefigure{S\arabic{figure}}

\section{Network architecture}\label{sec::network_architecture}
A fully-connected feed-forward network (FFNN) is a composition of $\mu \in \mathbb N_{>0}$ layers where the $l$th network layer, $1 \le l \le \mu$, applies the transformation (writing $[n] := \{1, \ldots, n\}$ for any $n \in \mathbb N_{>0}$)
\begin{eqnarray}\label{eq::architecture}
\begin{split}
    z_{i}^\pars{l} = \sum_{j}W_{ij}^\pars{l}y_{j}^\pars{l} + b_{i}^\pars{l}, \qquad y_{i}^\pars{l+1} = \phi(z_{i}^\pars{l}) \\
    \begin{cases}
        i \in [L], \, j \in [\alpha L], & l = 1, \\
        i, j \in [\alpha L],                   & l > 1,
    \end{cases}
\end{split}
\end{eqnarray}
from input $y^\pars{l}$ to output signal $y^\pars{l+1}.$
This is an affine transformation with weight matrix (or kernel) $W^\pars{l}$, bias $b^\pars{l},$ and a nonlinear activation function $\phi.$
The full set of network parameters, consisting of all weights and biases, is denoted by $\theta \in \mathbb{C}^{\Npar}.$
For the results presented in the main text, we have used the SELU activation function
\begin{align}
    \mathrm{SELU}(x) = \lambda \begin{cases}
        x, & x > 0, \\
        (a - 1) e^x, & x \le 0
    \end{cases}
\end{align}
(where $\lambda \approx 1.0507$ and $a \approx 1.6733$),
as implemented in \textsc{jax} \cite{Jax2018},
which helps circumvent the vanishing gradient problem of training deep networks \cite{Klambauer2017}.
Since in our case the network acts on complex-valued signals, we apply SELU separately to real and imaginary part of the input (\texttt{reim{\textunderscore}selu} in \textsc{netket} \cite{NetKet3}), giving the complex activation
\begin{align}
    \phi(z) = \mathrm{SELU}(\Re z) + i \, \mathrm{SELU}(\Im z).
\end{align}
Using the hyperbolic tangent as activation provides equivalent or (for deep networks) worse results, as we discuss below (\cref{sec:tanh}).
The log-probability amplitudes of the variational state are then obtained by performing the logsumexp operation on the output of the final layer, i.e.,
\begin{equation}
       \log \braket{x|\psi_{\theta}} = \log\left(\sum_{i=1}^{\alpha L} \exp(y_{i}^\pars{\mu + 1})\right),  \qquad y^\pars{1}_{i} = x_i,
\end{equation}
with $\{\ket{x} \mid x \in \{0,1\}^L \}$ denoting the computational basis.

\section{Training protocol}

\begin{figure}[tb]
\centering
\minipage{0.45\textwidth}
  \includegraphics[width=1\linewidth]{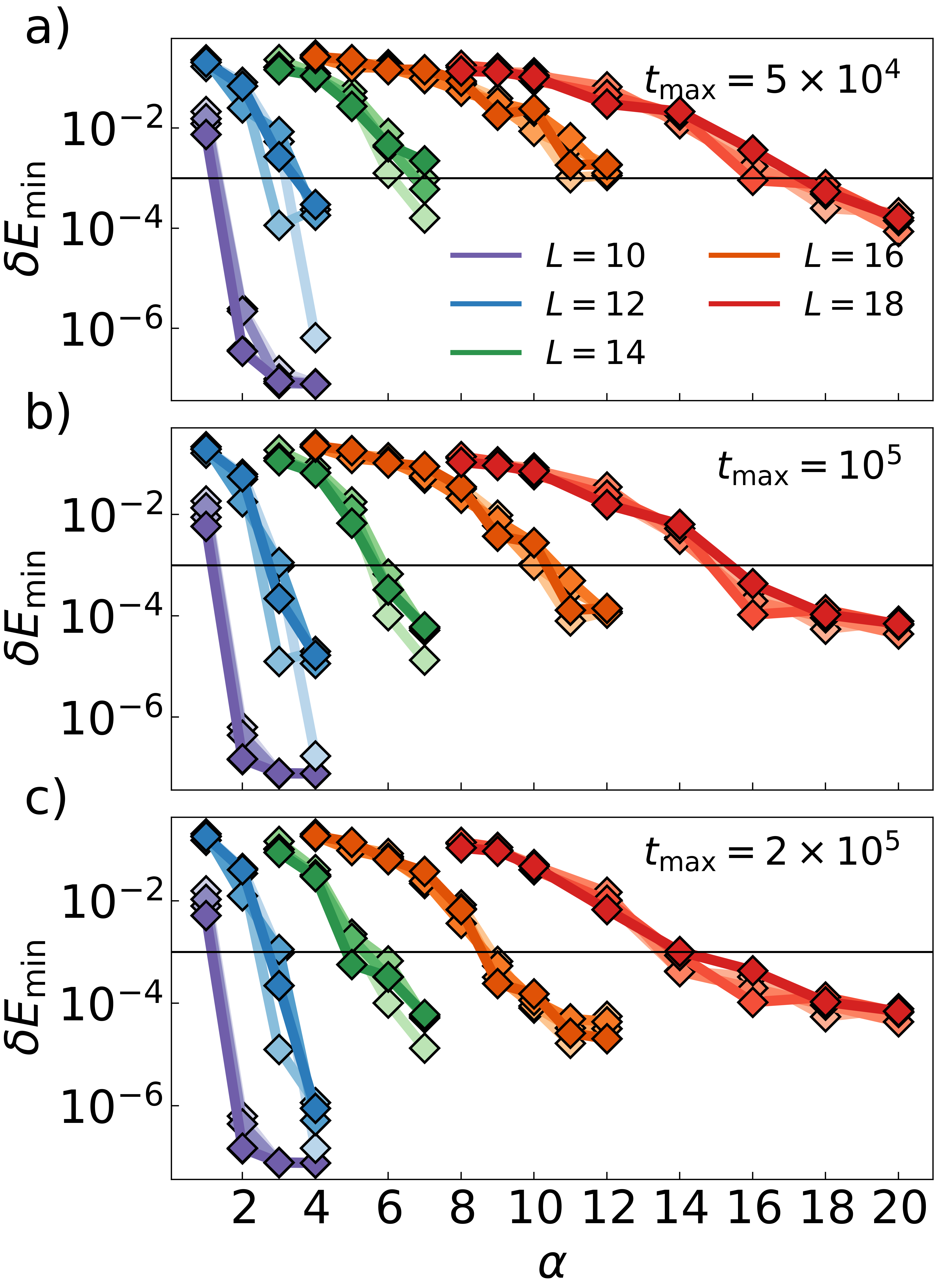}
\endminipage\hfill
\minipage{0.45\textwidth}
  \includegraphics[width=1\linewidth]{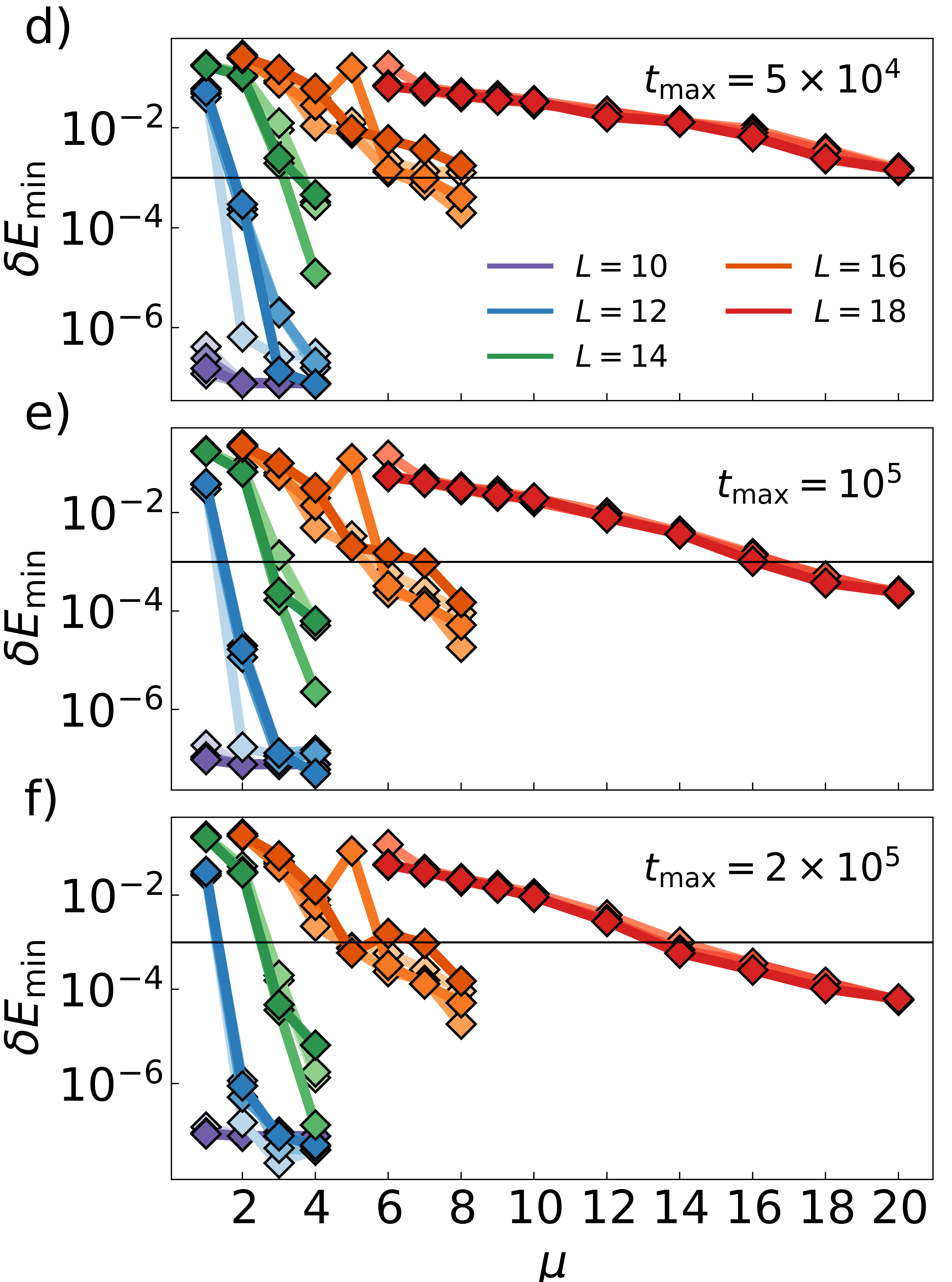}
\endminipage\hfill
\caption{Relative energy error $\delta E$ as function of [(a), (b), (c)] the network width $\alpha$ and of [(d), (e), (f)] the number of layer $\mu$ for several system sizes and random initializations after [(a), (d)] $5 \times 10^{4}$, [(b), (e)] $10^{5} $, and [(c), (f)] $2\times 10^{5}$ simulation steps, respectively.
Here we report the full dataset from which the average value plotted in the main text has been derived.
The color of each set of data points corresponds to the system size $L$ as indicated in the legend while different shades of the same color belong to independent realizations of the initial network weights.}
\label{fig:delta_vs_alpha}
\end{figure}

\subsection{Supervised learning}

In this work we considered only system sizes up to $L=18$, which are solvable via exact diagonalization (ED).
We solve the ED problem for a given realization of the model $\HSYK(J)$, finding the corresponding ground state $\ket{\psi_{\rm GS}(J)}$.
The key result of this work, the exponential scaling of the required network size to reach a specific error bound, is independent from the specific $J$ realization for the couplings of the SYK model.
This has been checked by performing the full analysis as described in the following paragraphs for two independent realizations of the random coupling matrix $J$.

Given the ED solution $\ket{\psi_{\rm GS}}$, it is possible to train the network to reproduce the correct probability amplitudes (up to norm and phase gauge freedoms) via supervised learning \cite{Jonsson2018}.
In practice, we have done so by optimizing the loss function 
\begin{equation}
    \delta O(\theta, J) =  1 - \left|\frac{\braket{\psi_{\theta}|\psi_{\rm GS}(J)}}{\braket{\psi_{\theta}|\psi_{\theta}}} \right|
\end{equation}
using the Adam optimizer \cite{Kingma2014}.
Another training scheme can be implemented by directly minimizing the expected energy $E(\theta, J) = \braket{\HSYK(J)}_{\ket{\psi_\theta}}$ of the system as the loss function.
For this loss function, it is possible to extend the training to system sizes beyond the reach of ED studies through variational Monte Carlo (VMC) sampling \cite{Carleo2017}, which is not possible in the SL scheme employed here due to its reliance on the knowledge of the full solution vector.
Since all our system sizes could be treated via full summation, we have not relied on VMC sampling in this work.

As our goal is to find the network size required to represent the SYK ground states at a given system size, we have compared training runs for both SL and energy minimization routines, finding the SL to be the best performing choice in the regime under consideration (compare \cref{sec:voe}).
Therefore, we have selected the SL results for presentation in the main text.
Our observations regarding the exponential scaling of required network size hold also for the variational energy optimization runs.

The optimization problem for a neural network is challenging on its own and there is unfortunately no general way to systematically identify the best choice of hyperparameters.
Here, we have chosen to study the scaling with regard to width $\alpha$ and depth $\mu$ of the network by performing two sets of runs: Varying the width for a two-layer network and varying the depth of a network of fixed width $\alpha = 4.$
Considering the trajectory of weights $\{\theta(t)\}_{t \in [t_{\rm max}]}$ (where $[t_{\rm max}] := \{1, \ldots, t_{\rm max}\} \subseteq \mathbb N$) over the sequence of optimization steps, we can the define the optimal energy relative error as $\delta E_{\rm min}(\theta, J) = {\rm min}_{t \in [t_{\rm max}]}E(\theta(t), J)$.
In \cref{fig:delta_vs_alpha} we report $\delta E_{\rm min}$ for SL protocols for three different number of total iterations $t_{\rm max}$ as function of $\alpha$.
All the data reported in this plot have been obtained with $\mu = 2$, keeping the learning rate of the Adam optimizer constant throughout all the simulation.
From this analysis we can make two relevant observations.
The first is that up to the network sizes simulated there is a monotonic improvement of the error with increasing width $\alpha$.
The second observation is that at each fixed number of simulation steps there is an exponential scaling of the $\alpha_{\rm min}$ required to bring $\delta{E_{\rm min}}$ below an arbitrary threshold ($10^{-3}$ in our example) as a function of the system size.
With the data reported in this paragraph, it cannot be ruled out that this exponential cost in network size is due to an exponential scaling of the number of steps required to bring the small $\alpha$ values to convergence.
However, through an analysis of the training curves it is possible to make a strong argument to identify the $\alpha_{\rm min}$ discussed in the main text as a quantity independent from the number of training steps.
This analysis is explained in the following section.

\begin{figure}[tb]
   \includegraphics[width=0.8\textwidth]{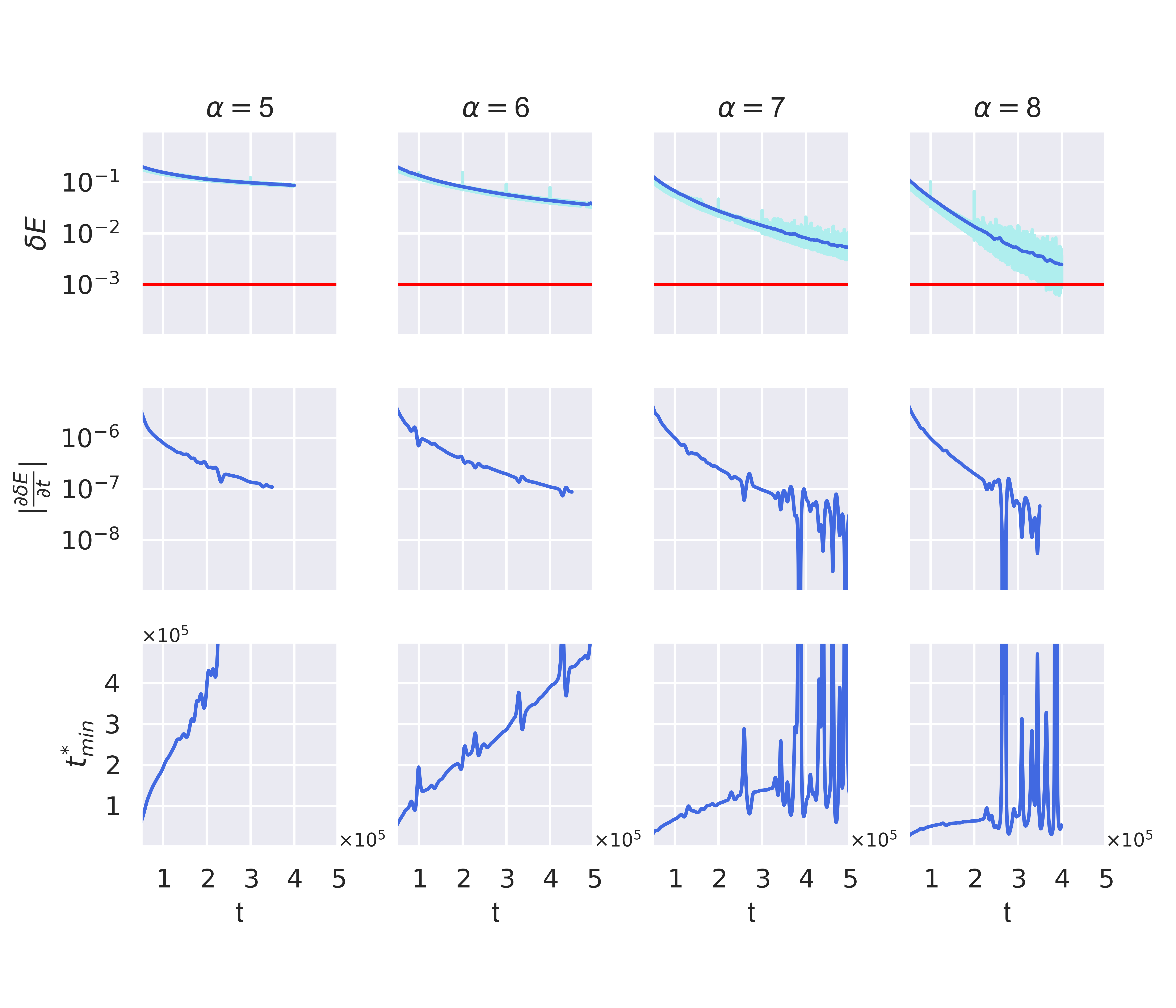}
\caption{(a)~Relative energy error $\delta E$ as function of the iteration step for shallow networks with $\mu = 2$ layers and width $\alpha$ as reported in the respective panels. The system size is $L = 16$. In light blue we report the raw data, in dark blue the result of a window smoothing filter applied to the raw data.
(b)~Absolute value of the numerical derivative $|\frac{\partial(\delta E)}{\partial t}|$. (c)~Estimated minimum number of steps required to convergence $t^{*}_{\rm min}$ as defined by \cref{t_estimator}. }
\label{fig:truncation_plot_energy}
\end{figure}

\subsection{Truncation scheme}
In the first row of \cref{fig:truncation_plot_energy}, we report $\delta E(\theta(t), J)$ for four different alpha values distributed around what we are going to define as the critical $\alpha_{\min}$.
The light blue line corresponds to the raw data obtained from the training process.
In order to remove the fluctuations around the moving average, we apply a flat window filter over the raw error, obtaining the profile corresponding to the dark blue line.
This running average error is smooth and monotonically decreasing.
Furthermore, the slope of these curves is also decreasing (in absolute value), implying a slowing rate of convergence of the optimization.
In order to conserve computational resources, we truncate the runs that are expected to converge to a $\delta E(\theta, J)$ above the energy threshold of $10^{-3}$ based on the following criterion:
Assuming that the absolute value of the slope is monotonically decreasing throughout all the training iterations, it is possible to estimate, at any step $t$, a lower bound for the number of steps required to reach the error threshold as
\begin{equation}\label{t_estimator}
    t^{*}_{\rm min}(t) = \frac{\delta E(\theta(t), J)- \delta E^{\rm threshold}}{|\partial_{t}\delta E(\theta(t), J)|}.
\end{equation}
Within the assumption of constant slope, the equality $\delta E(\theta(t + t^{*}_{\rm min}(t)), J) = \delta E^{\rm threshold}$ holds.
If now we define a control interval $\Delta t$ big enough to average out the noise around training curve, one can check if a candidate run is trending towards convergence below the error threshold based on the slope of $t^{*}_{\rm min}(t)$,
\begin{equation}
     \delta t^{*}_{\rm min}(t) = \frac{t^{*}_{\rm min}(t + \Delta t)- t^{*}_{\rm min }(t)}{\Delta t}.
\end{equation}
The runs that are expected to reach convergence are those that satisfy
\begin{equation}
\label{eq:criterion}
    \delta t^{*}_{\rm min}(t + \Delta t) < \delta t^{*}_{\rm min}(t).
\end{equation}
We simulated all runs for at least $t_{\rm max} = 2 \times 10^{5}$, for those runs that where still above the threshold for that step we used the criterion \eqref{eq:criterion} (with $\Delta t = 10^5$) in order to assess whether they can still be expected to reach the error threshold, based on their current rate of convergence.
The runs for which converge below the threshold was ruled out by our criterion were stopped in order to conserve computational resources, while we did perform additional blocks of $10^5$ steps for the remaining simulations, until the truncation criterion \eqref{eq:criterion} did apply or the error threshold was reached.
The resulting $\alpha_{\rm min}$ values reported in Fig.~4 of the main text are the smallest simulated value of $\alpha$ for which convergence below the threshold was reached.
The $\mu$ scaling analysis was performed in an equivalent way.

\begin{figure}[tb]
   \includegraphics{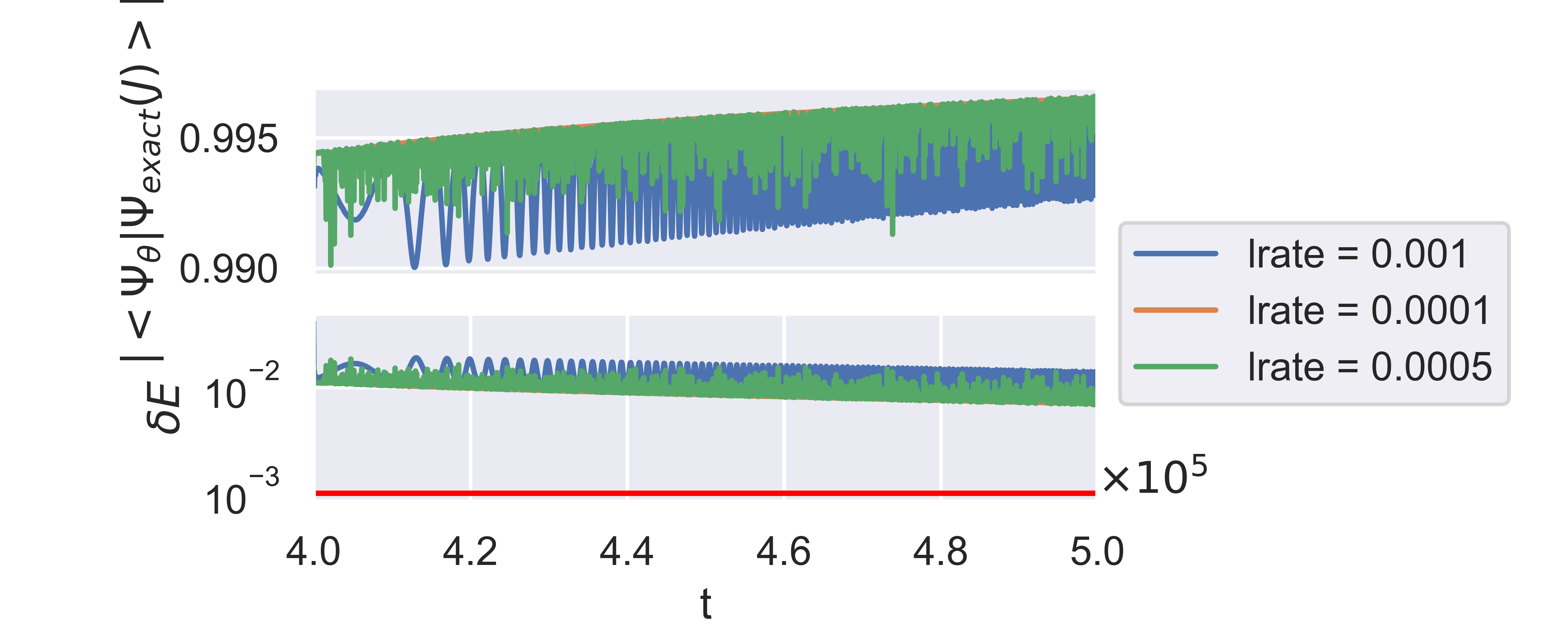}
\caption{(a) Overlap $\braket{\Psi_{\theta}|\Psi_{\rm exact} (J)}$ and (b) relative energy error $\delta E$ for a training protocol in the interval $t \in [4\times 10 ^{5}, 5 \times 10^{5}]$ steps. 
The same pre-trained network here it has been updated with the three different learning ratios reported in the legend.}\label{fig:training_hyper}
\end{figure}

\section{Overview of alternative training methods investigated}

In the previous paragraphs we have described the protocol and data analysis necessary to reproduce the results that we discussed in the main text.
Here we are showing alternatives to the training scheme and hyperparameters.
The alternatives which we have tested achieved either worse accuracy or no significant improvement compared to the results we selected for the main text.

\subsection{Different learning rate}
We have used the Adam optimizer using a constant learning rate throughout the simulations reported in the main text. 
One advantage of Adam with respect to SGD is that it automatically adapts the step size based on the history of loss gradients.
Still, Adam has to be initialized with a specific learning rate and this can potentially influence the effectiveness of the training.
While determining the optimal learning rate for our simulations, we have also tried schemes involving an update of the learning rate during the simulation.
We report an example of an unconverged training curve in \cref{fig:training_hyper} for which, after $4\cdot 10^{5}$ steps at a fixed learning rate, we performed further $10^{5}$ steps with three different adjusted learning rates.
These changes of learning rate did not result in an improvement of training performance.

\subsection{Variational energy optimization}
\label{sec:voe}
\begin{figure}[tb]
   \includegraphics{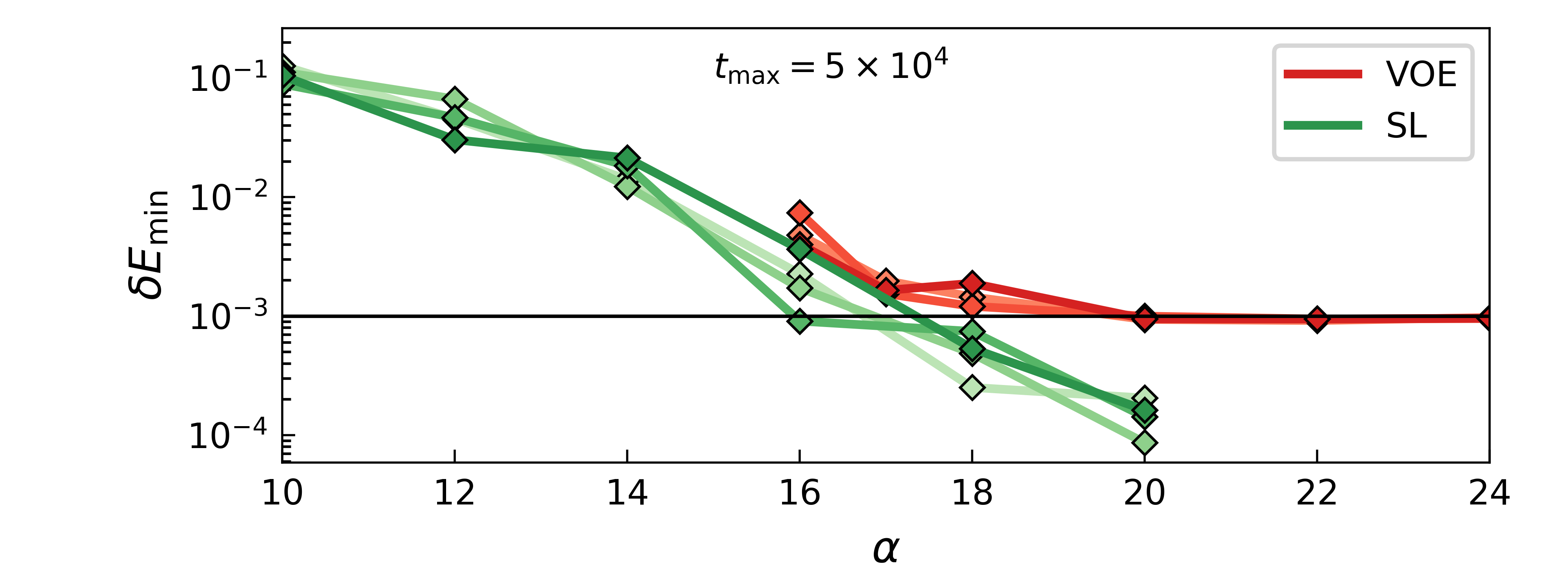}
\caption{Minimum relative energy error $\delta E_{\rm min}$ obtained using the SL (green) and a VOE (red) loss functions using $t_{\rm max} = 5 \times 10^{4}$ steps as function of the number of network width $\alpha$.
For the data reported in this plot, we considered system size $L = 18$ and a fixed $\mu = 2$}
\label{fig:SL_vs_VOE}
\end{figure}

Beside the SL protocol described above, we have also considered directly optimizing the variational energy
\begin{equation}
\label{eq:voe}
    E(\theta, J) = \frac{\braket{\psi_{\theta}|\HSYK(J)|\psi_{\theta}}}{\braket{\psi_{\theta}|\psi_{\theta}}}.
\end{equation}
For disambiguation, we refer to this loss function as the variationally optimized energy (VOE).
Note that this loss does not involve the ED solution of the model and can be applied to system sizes beyond those limits, if it is used together with MC sampling of the quantum states \cite{Carleo2017}.
However, as we are interested in comparing the performance of the SL and VOE protocols in the ED regime, we only use the loss \eqref{eq:voe} obtained by full summation in the following.
Within the parameter range tested, the SL protocol achieved systematically lower errors than the VOE loss.
As an example of this, we show in \cref{fig:SL_vs_VOE} the resulting $\delta E_{\rm min}$ for a given system size ($L = 18$) and a fixed number of iteration steps. 
Although the trend is similar between the two curves, the SL reaches the convergence threshold at lower $\alpha$ values.

\begin{figure}[tb]
   \includegraphics{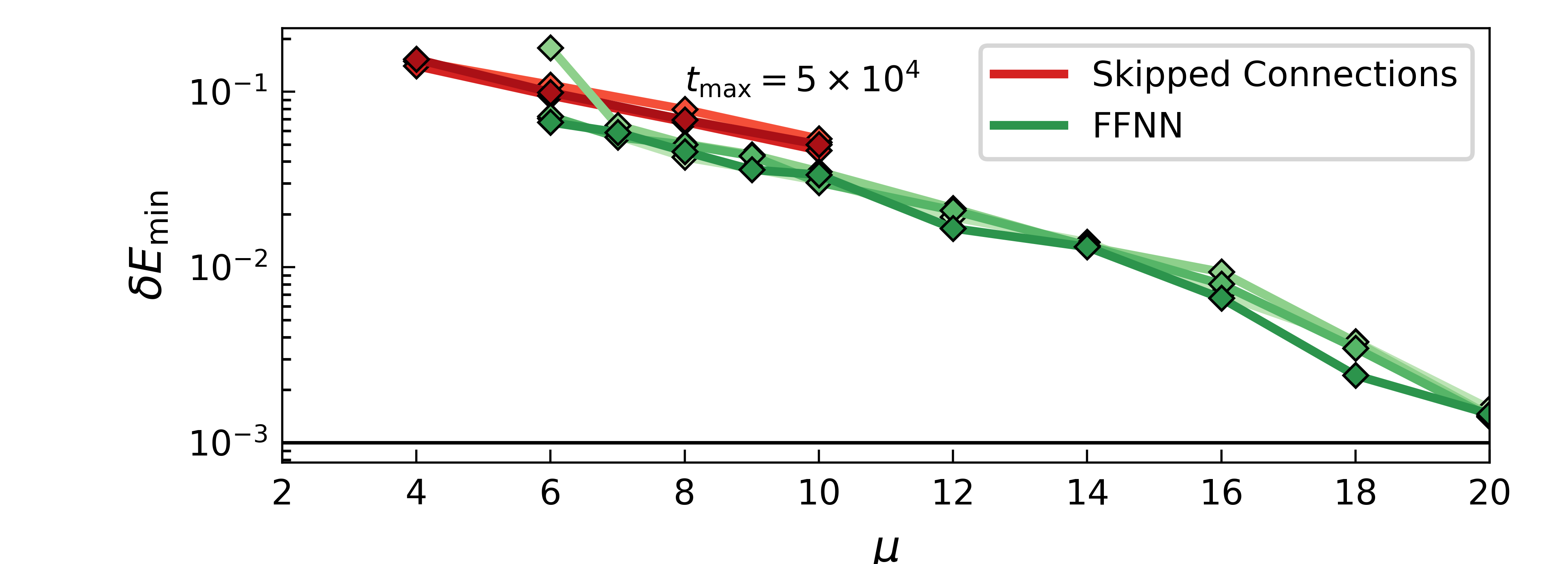}
\caption{Minimum relative energy error $\delta E_{\rm min}$ obtained during a SL training protocol of a deep FFNN (green) and of a Network with Skipped connections (red) with a number of steps set to $t_{\rm max} = 5 \times 10^{4}$ as function of the number of network laters $\mu$.
For the data reported in this plot we considered system size $L = 18$ and a fixed $\alpha = 4$}
\label{fig:skip}
\end{figure}

\subsection{Deep networks with skip connections}
In the main part of this work, we presented an extensive discussion of the results obtained when training a FFNN with the architecture defined in \cref{sec::network_architecture}.
Among the two different hyperparameter dependencies discussed there, the number of layers $\mu$ scaling requires some extra attention.
It is known that while adding more layers enhances the networks expressive capability, at the same time it can make the training more difficult \cite{He2016}.
One common remedy to this is the addition of skip connections between the layers of the neural network \cite{Li2018b}.
We report in \cref{fig:skip} a comparison between the training at a fixed number of iteration steps $t_{\rm max} = 5 \cdot 10^{4}$ between the FFNN discussed in the main text and a fully-connected network with skip connections.
Specifically, our network is divided into $n_B$ blocks, each containing $\ell$ layers.
The total number of layers is thus $\mu = n_B \ell.$
After the application of each block (i.e., after applying $\ell$ fully-connected layers), the content of the original input layer (with the first affine transformation applied) is added to the output before it is passed to the following block.
For the SL simulation runs considered (\cref{fig:skip}), this architecture does not give an improvement in energy error when compared to a simple FFNN.
We have found improved convergence (compared to the simple FFNN) of the skip connection networks in some runs using the VOE loss.
However, in those cases the number of layers already exceeded the critical $\mu$ (around which the network with skip connections still performed worse), so no improvement with regard to compression was obtained this way.

\begin{figure}[tb]
   \includegraphics[width=0.8\textwidth]{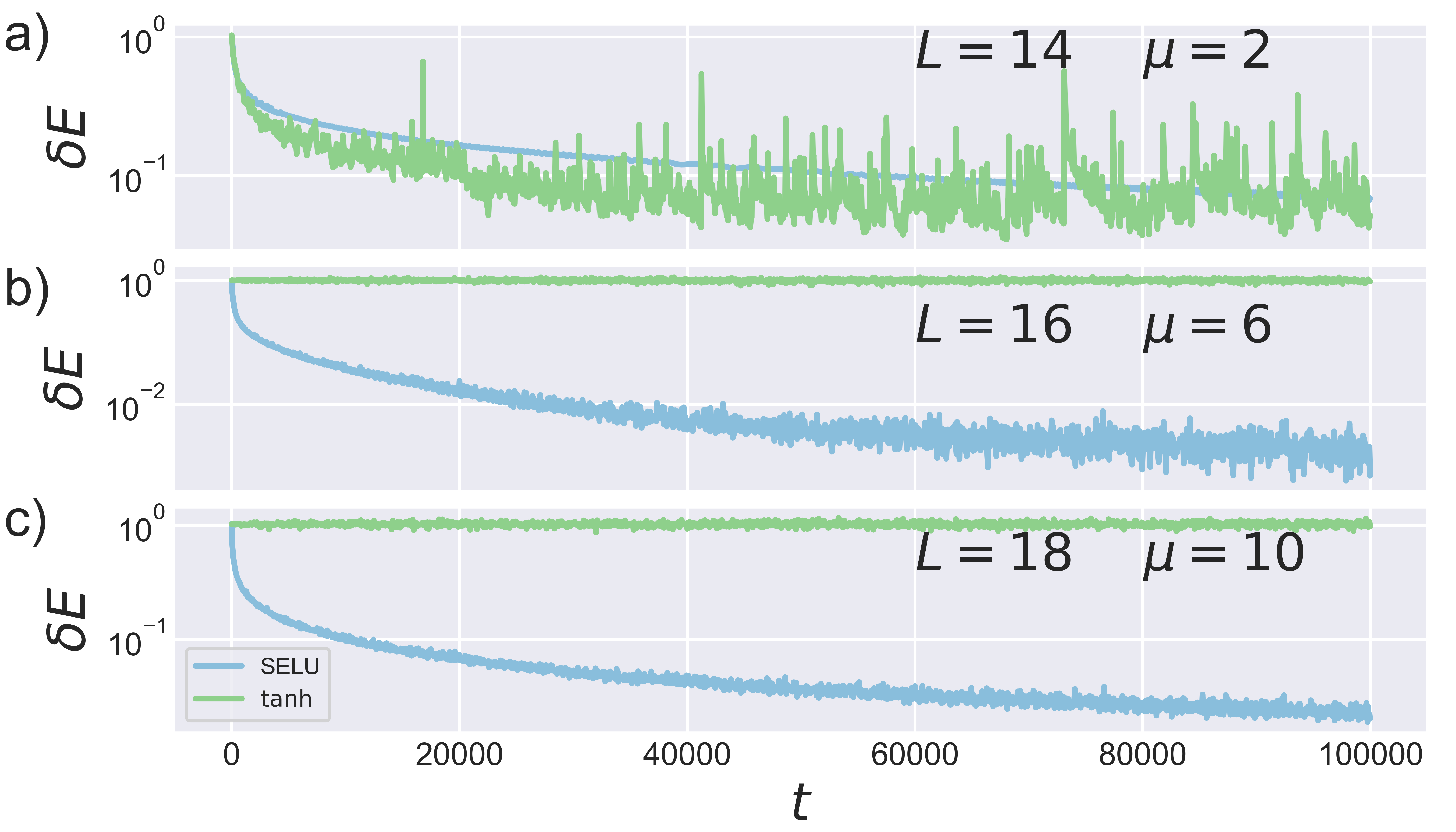}
\caption{Relative energy error $\delta E$ as function of the iteration step $t$ using the SL training protocol with SELU (light blue) and $\tanh$ (light green) activation functions.
The figure shows the data for (a) $L = 14$, $\mu = 2$; (b) $L = 16$, $\mu = 6$; and (c) $L = 18$, $\mu = 10$.
}
\label{fig:tanh_comparison}
\end{figure}

\subsection{Hyperbolic tangent activation function}
\label{sec:tanh}
As already stated, for all the data discussed in the main part, the FFNN implemented where following the definition presented in Eq.~\eqref{eq::architecture} using SELU as activation function $\phi$.
While in principle there are many viable choices with regard to which specific activation function utilizing, we opted to restrict our studies to a particular activation in order to investigate extensively the dependence of the representability power as function of the network hyperparameters, keeping a manageable number of simulations to perform.
Thus the need to choose an activation function able to achieve reliable performance when dealing with both deep and shallow networks.
We present here a comparison of the performances achieved by the same training protocol and network architecture while using another common activation function, the hyperbolic tangent.
In \cref{fig:tanh_comparison}, we present training profiles for the relative energy error $\delta E$ as function of the iteration step $t$.
We show this for both a network with SELU and a network with $\tanh$ activation function for several numbers of layers $\mu$.
As one could expect from the vanishing gradient problem that effects $tanh$, we observe that also for the network sizes relevant to our problem, above a certain number of layers the $\tanh$ activated networks are not able to improve the energy error anymore.
At the same time SELU displayed a monotonic improvement of the energy error with the number of layers (up to scales relevant for our specific problem) and thus is an effective choice (compare \cref{fig:delta_vs_alpha}).

\begin{figure}[tb]
   \includegraphics{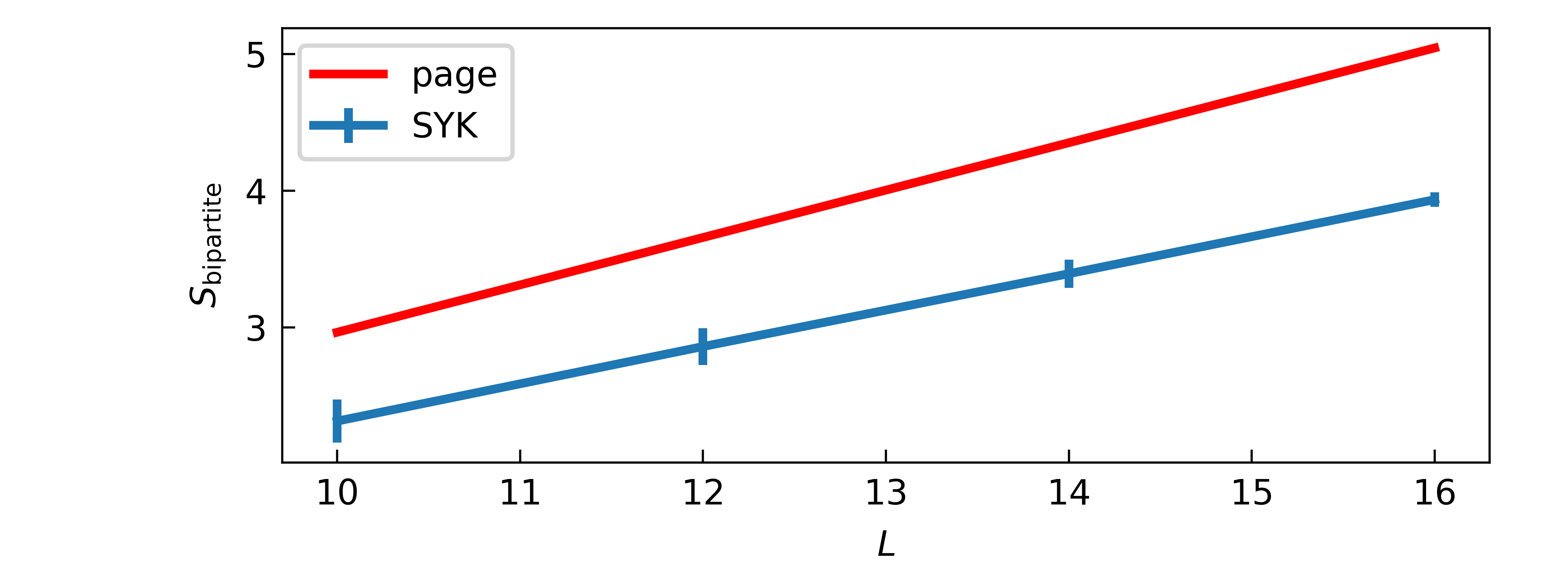}
\caption{Bipartite entanglement entropy averaged over 4 independent realizations of the SYK model (error bar shows the standard deviation) compared to bipartite entanglement of a random state given by the Page value (red).}
\label{fig:entanglement}
\end{figure}

\section{SYK bipartite entanglement}
Following the standard definition, the bipartite entanglement of a density matrix $\rho$ corresponds to
\begin{equation}
    S_{\rm bipartite} = - {\rm Tr}\left[ \rho_{A}\log \left(\rho_{A}\right)\right] = - {\rm Tr}\left[ \rho_{B}\log \left(\rho_{B}\right)\right]
\end{equation}
where $\rho_{A(B)} = {\rm Tr}_{B(A)}\left[\rho \right]$ is the reduced density matrix obtained from the density matrix $\rho$ and tracing over the degrees of the subpartition B(A) of the total hilbert space, in the special case of ${\rm dim}(A)={\rm dim}(B)$. 
We compare the $S_{\rm bipartite}$ scaling for four independent finite size SYK model realizations together with the Page value \cite{Page1993}, that quantifies the entanglement for a pure random state.
In the special case case of bipartite entanglement, the page value is
\begin{equation}
    S_{\rm page} = N\log(2)-\frac{1}{2}.
\end{equation}
As one can see in \cref{fig:entanglement}, the entanglement scaling of the SYK model follows a volume law scaling that does not saturate the page value, showing that the SYK states exhibit structure beyond a pure random state.

\begin{figure}[tb]
   \includegraphics[width=\textwidth]{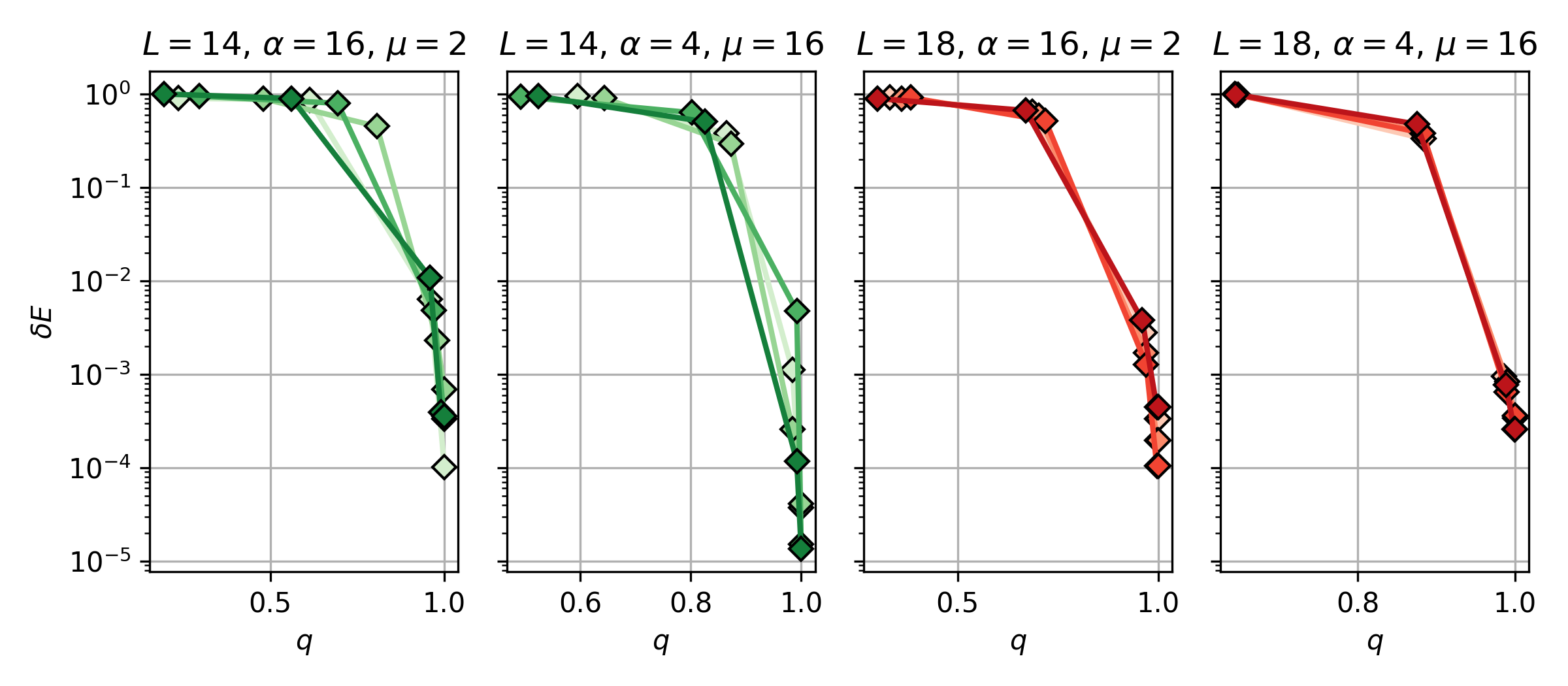}
\caption{
    Relative energy error as a function of the fraction of retained singular values $q$
    for four different choices of hyperparameters indicated in the title of each panel.
    The color corresponds to different network initializations as in \cref{fig:delta_vs_alpha}.
}
\label{fig:svd_compression}
\end{figure}

\section{Network compression}

Trained neural networks can be amenable to compression in order to reduce the total number of parameters.
This is usually done for the purpose of reducing memory and computational requirements in order to run a network on a lower powered device.
Here, we have explored a compression scheme based on a lower-rank approximation of the weight matrices in each layer \cite{Xue2013} in order to gauge whether any potential redundancy in the learned variational parameters can be identified this way.
This is done by performing a singular value decomposition (SVD)
\begin{align}
    W^\pars{l} = U^\pars{l} S^\pars{l} V^\pars{l}
\end{align}
where $U^\pars{l}$, $V^\pars{l}$ are unitary matrices and $S^\pars{l} = \operatorname{diag}(\sigma_1^\pars{l}, \ldots, \sigma_M^\pars{l})$ is the matrix of singular values $\sigma_1^\pars{l} \ge \ldots \ge \sigma_M^\pars{l} \ge 0$.
We truncate this spectrum by discarding all singular values below a threshold $\lambda$ relative to the largest singular value, i.e., based on the criterion $\lfrac{\sigma_i^\pars{l}}{\sigma_1^\pars{l}} < \lambda.$
For simplicity, we chose $\lambda$ uniformly for each layer.
As a measure of the approximation error, we report in \cref{fig:svd_compression} the relative energy error after truncation as a function of the fraction
\begin{align}
    q = \frac{\text{singular values retained}}{\text{total singular values}}
\end{align}
of singular values retained over all weight matrices of the network for different system sizes and choices of hyperparameters.
This data shows that already a limited amount of truncation ($0.95 \le q < 1.0$) results in a significant increase in energy error.
We therefore conclude that at uniform SVD truncation of the weight matrices does not reveal redundancy in our learned SYK ground states.

\bibliography{bibliography.bib}